\documentclass[reprint,prx,aps,superscriptaddress,amsmath,amssymb]{revtex4-2}

\usepackage{graphicx}
\usepackage{dcolumn}
\usepackage{bm}
\usepackage{color}
\usepackage[dvipsnames]{xcolor}
    \definecolor{Sergio}{rgb}{0, 0.25, 1}
    \definecolor{lplajaOrange}{RGB}{255,100,0}
    \definecolor{carlos}{RGB}{150,0,150}
\usepackage{placeins}

\usepackage[normalem]{ulem}
\usepackage{xparse}
\usepackage{soul} 

\soulregister\cite7
\soulregister\ref7
\soulregister\pageref7
\soulregister\footnote7

\usepackage[linkcolor=black,urlcolor=black,citecolor=black,colorlinks=true]{hyperref}

\NewDocumentCommand{\lplaja}{s m}{%
  \begingroup
  \color{lplajaOrange}%
  \IfBooleanTF{#1}{\st{#2}}{#2}%
  \endgroup
}

\NewDocumentCommand{\chg}{s m}{%
  \begingroup
  \color{carlos}%
  \IfBooleanTF{#1}{\st{#2}}{#2}%
  \endgroup
}

\begin{document}

\title{Attosecond pulse trains from graphene via macroscopic phase-matching in high harmonic generation}

\author{Sergio Mart\'in-Domene}
\email{sergiomardom@usal.es}
\affiliation{Grupo de Investigaci\'on en Aplicaciones del L\'aser y Fot\'onica, Departamento de Física Aplicada, Universidad de Salamanca, E-37008, Salamanca, Spain}
\affiliation{Unidad de Excelencia en Luz y Materia Estructuradas (LUMES), Universidad de Salamanca, Salamanca, Spain \looseness=-1 }

\author{Luis Plaja}
\affiliation{Grupo de Investigaci\'on en Aplicaciones del L\'aser y Fot\'onica, Departamento de Física Aplicada, Universidad de Salamanca, E-37008, Salamanca, Spain}
\affiliation{Unidad de Excelencia en Luz y Materia Estructuradas (LUMES), Universidad de Salamanca, Salamanca, Spain \looseness=-1 }

\author{Carlos Hern\'andez-García}
\affiliation{Grupo de Investigaci\'on en Aplicaciones del L\'aser y Fot\'onica, Departamento de Física Aplicada, Universidad de Salamanca, E-37008, Salamanca, Spain}
\affiliation{Unidad de Excelencia en Luz y Materia Estructuradas (LUMES), Universidad de Salamanca, Salamanca, Spain \looseness=-1 }

\begin{abstract}
Attosecond pulse generation in solids remains challenging due to the complex phase structure arising from multiple electronic pathways in high harmonic generation (HHG). Here, we identify the macroscopic conditions that enable the synthesis of attosecond pulse trains in graphene via HHG. Using numerical simulations that combine microscopic and macroscopic HHG in single-layer graphene, we show that harmonic contributions associated with different emission times acquire distinct far-field beam-divergence properties, analogous to the short- and long-trajectory contributions in gas-phase HHG. Our results identify how transverse phase-matching can be engineered in solid-state HHG to select short-time or long-time electronic contributions through proper tailoring of the driving-field waveform. 
In particular, the macroscopic suppression of long-time contributions associated with delayed electron-hole recombinations or imperfect recollisions---
usually removed in semiconductor Bloch equation calculations by introducing artificial decoherence times---
leads to clean, positively chirped, attosecond pulse trains with a temporal quality comparable to that achieved in gas-phase HHG. These results establish a general framework for controlling attosecond emission in solid-state HHG and provide a route toward compact solid-state attosecond sources.

\end{abstract}

\maketitle


\section{Introduction}

The manipulation of electronic dynamics in atoms, molecules and solids in their natural timescales has become possible with the generation of light pulses reaching durations at the attosecond timescale. These pulses open access to ultrafast processes underlying  thermal transport \cite{Siemens2010}, elastic scattering \cite{Karl2018}, many-body effects \cite{Lee2024}, photoelectron-state formation \cite{Laurell2025}, and molecular dynamics \cite{Calegari2014, Severino2024}, among others \cite{BorregoVarillas2022}. Reaching the attosecond regime requires three essential ingredients: (i) sufficiently high-frequencies, in the extreme-ultraviolet (EUV) or beyond, ensuring the laser cycle lies in the sub-femtosecond regime; (ii) broad spectral bandwidth to support ultrashort pulses; and (iii) a well-defined spectral phase relationship---phase-locking---accross the emitted spectrum. These requirements are naturally met in high-order harmonic generation (HHG) driven in gases, a highly nonlinear process discovered in the late 80’s \cite{McPherson1987, Ferray1988}, which laid the foundation of attosecond science.

In gas-phase HHG, an intense femtosecond infrared (IR) laser interacts with an atomic or molecular target, inducing electron tunnel ionization. The liberated electronic wavepacket is accelerated by the oscillating laser field, and recollides with the parent ion, emitting high-frequency radiation upon recombination \cite{Corkum1993, Schafer1993}. The radiation is emitted as a comb of high-order harmonics that can sustain a train of attosecond pulses \cite{Farkas1992, Paul2001}. Driving the process with few-cycle pulses can confine the emission to an isolated attosecond pulse \cite{Christov1997, Hentschel2001, Sola2006}. The emission of attosecond pulses is enabled by the intrinsically regular and well-defined spectral phase of HHG in gases. The quantum phase associated with the electron paths contributing to recombination---known as short and long trajectory contributions---are imprinted on the spectral phase \cite{Lewenstein1995, Mairesse2003}. Proper phase-matching conditions lead to the constructive interference of the short trajectory contributions when integrated over the macroscopic target \cite{Salieres1995, Gaarde2008_PM}, resulting in trains of positively chirped subfemtosecod pulses, that can be compressed to durations down to tens of attoseconds \cite{Li2017, Gaumnitz2017, Chien2024, Ardana-Lamas2025}.

Despite its technical maturity, gas-phase HHG still faces practical limitations in conversion efficiency and scalability towards the development of compact attosecond platforms. This has motivated efforts to extend HHG to condensed matter targets, where strong-field-driven currents potentially enable brighter, chip-scale EUV ultrafast sources. In addition, solid-state harmonics have been demonstrated as a powerful probe for ultrafast phenomena in quantum materials \cite{Ghimire2019}, in applications as EUV spectroscopy \cite{Luu2015, GarciaCabrera2024}, and the observation of electron-hole correlations in two-dimensional materials \cite{Heide2022} and light-driven valleytronic responses \cite{Mitra2024, Tyulnev2024}, among others.

HHG in solid-state targets can be understood as the result of the field-driven electron excursions in the reciprocal lattice. Harmonic emission may originate from intraband electron dynamics along non-parabolic energy bands, as well as from interband electron-hole recombination at the instants when the associated trajectories recollide in real space. In finite-gap materials, the electron-hole pairs are generated from tunnel excitation from the valence band (VB) to the conduction band (CB) at the gap minima, at the instants when the driving field amplitude reaches its maxima. 
In contrast, for the case of Dirac materials, the promotion of a VB electron to the CB follows from the non-adiabatic crossing of the electronic excursion near the Dirac points, at any time during the interaction. The presence of Dirac points, therefore, leads to more intricated quantum pathways contributing to each harmonic when compared with finite-gap materials, where short and long trajectories analogous to the case of gases can be identified \cite{Vampa2015, Soonyoung2022}. In the general case, the electron-hole recombination leads to the emission of high-order harmonics, with a photon energy corresponding to the valence-to-conduction band gap at the instant of recombination \cite{Ndabashimiye2016,Vampa2014,Floss2018,Higuchi2014,Luu2016}. 
In this context, solid-state HHG has been reported in a variety of finite-gap solids such as ZnO, $\mathrm{MoS}_2$, ZnSe, GaSe or $\mathrm{SiO}_2$ \cite{Ghimire2011,Liu2017,Luu2015,Hohenleutner2015,Lanin2017,Schubert2014}, and Dirac materials as single-layer graphene (SLG) \cite{Yoshikawa2017,Zurron2018} or topological insulators \cite{Ghimire2021_HHG_topological_insulator}.

Generating attosecond pulses from solid emitters also remains substantially more challenging than in the atomic case. Theoretically, recent studies have proposed the generation of isolated attosecond bursts in solids based on highly structured drivers in the time domain as two-color pulses \cite{Nourbakhsh2021,ZhaopinChen2025,Venkat2026} or more engineered tailored pulses \cite{GefeiLi2026}. Experimentally, the use of tailored light transients has enabled the production of bursts at around 24~eV in bulk solids with durations on the order of 500~attoseconds, attributed to intraband currents \cite{Garg2016}. Yet, a broadly applicable framework for attosecond synthesis in solids is still lacking, and theoretical reconstructions of solid-state attosecond waveforms remain much less regular than their gas-phase counterparts.

In contrast to gases, the condensed-matter response involves richer carrier dynamics, so that an harmonic order results from the contributions of more---and more complex--- electron pathways, which makes the spectral phase harder to regularize and control. This complexity has been simplified theoretically by introducing finite dephasing times \cite{Vampa2014,ZhaopinChen2025,Venkat2026}. More recently, however, several studies have suggested that an effective selection of quantum pathways may emerge naturally once macroscopic propagation and phase-matching effects are taken into account \cite{Mette.B.Gaarde2018_spatio_temporal_filtering, Boyero2021_transverse_phase_matching, G.G.Brown2023_part1, G.G.Brown2023_part2}. 
A complete description of HHG must account for both the coherent integration of all the local emissions within the target at the near field---due to the high sensitivity of the local harmonic field to the driving-field properties---and their propagation towards a far-field detector. Controlling these parameters in HHG in gases has allowed to optimize phase-matching not only to enhance the harmonic yield \cite{Gaarde2008_PM}, but also to select short-trajectory contributions \cite{Salieres1995}, to extend the maximum photon energy towards the soft X-rays \cite{Rundquist1997, Popmintchev2012}, and to shape the spatiotemporal structure of the emission, including the generation of isolated attosecond pulses with multicycle drivers \cite{Ferrari2010,Chen2014,Vismarra2024}. For solid targets, macroscopic integration shows the possibility to engineering  harmonic wavefronts via the intensity, phase, and polarization of the driver. However, while macroscopic propagation effects are well studied in gases, their detailed exploration in solids has only been initiated recently. Up to now, phase-matching studies have been focused in cleaning the HHG spectrum \cite{Floss2018,Boyero2021_transverse_phase_matching,G.G.Brown2023_part1, G.G.Brown2023_part2}, and increasing the harmonic yield \cite{Chenjun2025_TPM, Liang2024_BWPM}, but its influence in the generation of attosecond pulse trains remains unexplored.



In this article, we identify the macroscopic conditions under which attosecond pulse trains can be generated in graphene. Our numerical simulations of macroscopic HHG in SLG show that appropriate engineering of the driving field waveform enables the selection of harmonic contributions with sufficiently regular phase relationship to enable the synthesization of attosecond pulse trains upon propagation to the far field. In other words, we explore the transverse phase-matching conditions that allow the selection of electronic pathways that lead to the generation of regular attosecond pulse trains. We identify the presence of two dominant electronic contributions: short- and long-time recombination events---analogous to gas-phase HHG---, giving rise to harmonic beams with distinct divergence properties. These contributions are associated with emissions from the leading/trailing parts of the driving pulse (with corresponding blue-/red-shifted frequency contributions in each harmonic), the latter including so-called imperfect recollisions that introduce more complex phase structures and hinder attosecond pulse synthesization. We show that the different focusing characteristics of these contributions can be exploited to disentangle them in the far field. In particular, short-time recollision contributions lead to a well-defined far-field attosecond pulse train, in contrast to the intrinsically complex emission at the microscopic level. Analogous to trajectory selection in gas-phase HHG, the optimal focusing of the short-time recollision contributions---which can be achieved placing the SLG before the focus position---leads to a substantial improvement in the temporal coherence of the emitted pulses. Our results establish a pathway toward controlled attosecond pulse generation in solids, achieving temporal quality comparable to that of gas-phase HHG.

\section{Methods}

\begin{figure*}[t]
\centering
\includegraphics[width=\textwidth]{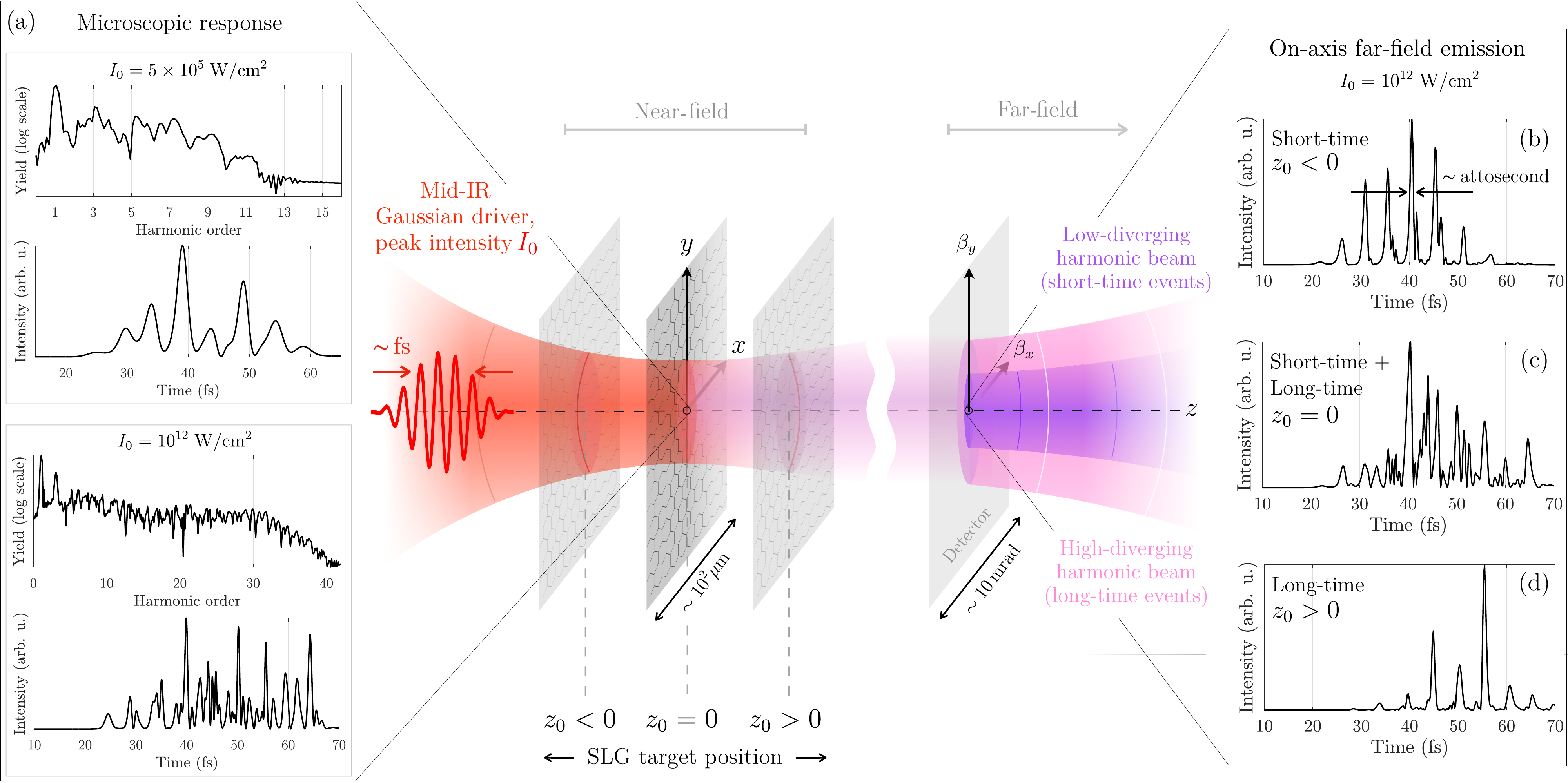}
\caption{Scheme for generating solid-state attosecond pulse trains. An intense mid-IR Gaussian driving beam is focused into a SLG sheet placed at a distance $z_0$ from the focus ($z=0$). The HHG emission is recorded in the far field, where different frequency components of the harmonic beam split in two main contributions with low and high divergence, associated to short- and long-time recombination events of the electron-hole pairs, respectively. The harmonic wavefront curvature depends strongly on $z_0$, and three representative cases are shown: $z_0 = -z_R, \ 0, \ +z_R$; $z_R$ being the Rayleigh range. (a) Microscopic HHG spectra and corresponding temporal emission in SLG driven by a linearly polarized, mid-IR 3~$\mu$m-wavelength, $28.8$~fs FWHM laser pulse with peak intensity $I_0 = 5 \times 10^{10}$~$\mathrm{W}/\mathrm{cm}^2$ (top) and $I_0 = 10^{12}$~$\mathrm{W}/\mathrm{cm}^2$ (bottom). The microscopic higher-order harmonic emission exhibits irregular temporal waveforms.  
(b-d) On-axis far-field temporal emission of the higher-order harmonics for the high intensity case $I_0 = 10^{12}$~$\mathrm{W}/\mathrm{cm}^2$, with (b) $z_0 = -z_R $, (c) $z_0 = 0$ and (d) $z_0= +z_R$. A clean attosecond pulse train is synthesized in the far-field from the low-divergence harmonic beam when the SLG sheet is placed before the focus. Depending on the target position, short- or long-time events are selected due to the macroscopic phase-matching.
}
\label{HHG_scheme}
\end{figure*}


The configuration for solid-state attosecond pulse generation is shown in Fig.~\ref{HHG_scheme}. We selected SLG as the target because, owing to its two-dimensional nature, longitudinal phase-matching effects can be neglected. This allows us to isolate transverse phase-matching effects \cite{Boyero2021_transverse_phase_matching}, namely the coherent buildup of harmonic emission across an effectively infinitesimally thin transverse target. In gas-phase HHG, transverse phase-matching is also commonly associated with the role of the spatially dependent waveform of the driving laser pulse \cite{hernandez-garcia2013,Quintard2019,wikmark2019spatiotemporal}. A mid-infrared (mid-IR), femtosecond, linearly polarized-Gaussian beam is focused onto the SLG sample, and the emission from the transverse layer is propagated to a far-field detector, where both the high-order harmonics and their temporal structure are characterized. We demonstrate that placing the SLG before the driving beam focus enables the generation of a regular attosecond pulse train suitable for experiemntal use, see Fig.~\ref{HHG_scheme}(b). Thus, the theoretical description requires two approaches: computing the microscopic HHG response of SLG, and integrating the macroscopic signal including transverse phase-matching and the propagation of the harmonics to the far field.

To compute the microscopic non-linear response of SLG \cite{book_graphene,Gaarde2022_tutorial}, we consider an effective two-band tight-binding model under the nearest-neighbours approximation. The dispersion relation for both VB ($-$) and CB ($+$) is $\epsilon_\pm (\mathbf{k}) = \pm \gamma_0 |f(\mathbf{k})|$, in terms of the hopping energy $\gamma_0 = 2.97$~eV and the complex geometric form factor $f(\mathbf{k})$. The complete wavefunction in reciprocal space of the Bloch electrons in the periodic lattice, $| \psi_{\mathbf{k}} \rangle$, can be expressed as a linear combination of the bands eigenstates $| \phi^{\pm}_{\mathbf{k}} \rangle$ obtained from the diagonalization of the tight-binding Hamiltonian $\mathcal{H}_0$. The density matrix operator of the system is then obtained as $\rho_{\mathbf{k}} = | \psi_{\mathbf{k}} \rangle \langle \psi_{\mathbf{k}} |$. Diagonal (real) terms $\rho_{-}$ and $\rho_{+}$ correspond to the VB and CB populations respectively, while off-diagonal (complex) terms $\rho_\pm = \rho_\mp^*$ are the microscopic polarizations or coherences.

When interacting with an external field $\mathbf{E}(t)$, the dipole-type interaction is introduced in the length gauge through the time-dependent Hamiltonian $\mathcal{H}(t) = \mathcal{H}_0 - q_e \hat{\mathbf{r}} \cdot \mathbf{E}(t)$, where $q_e$ is the electron charge and $\hat{\mathbf{r}}$ is the position operator. In this picture, the external field forces charge carriers oscillations in reciprocal space according to the acceleration theorem $\hbar \dot{\mathbf{k}} = q_e \mathbf{E}(t)$. Resorting to the Houston basis allows to further simplify the equations by replacing the crystal momentum by the kinetic quasimomentum $\hbar \boldsymbol{\kappa}_t = \hbar \mathbf{k} - q_e \mathbf{A}(t) / c$ in terms of the field vector potential $\mathbf{E}(t) = - \dot{\mathbf{A}}(t) / c$ and the speed of light $c$. 

The dynamics of the density matrix is captured by the semiconductor Bloch equations (SBE) which allows to include phenomenologically relaxation or decoherence effects in the optical response through a characteristic time $T_2$,
\begin{widetext}
\begin{equation}\label{SBE_bands_base}
    \left\{
    \begin{array}{lcl}
    \dot{\rho}_{+}(\boldsymbol{\kappa}_t,t) = & - \dot{\rho}_{-}(\boldsymbol{\kappa}_t,t) & = i \chi_{\pm}(\boldsymbol{\kappa}_t,t) \big[ \rho_{\mp}(\boldsymbol{\kappa}_t,t) - \rho_{\pm}(\boldsymbol{\kappa}_t,t) \big] , 
    \\
    \\
    \dot{\rho}_{\pm}(\boldsymbol{\kappa}_t,t) = & \dot{\rho}_{\mp}(\boldsymbol{\kappa}_t,t)^* & = -i \omega_g(\boldsymbol{\kappa}_t) \rho_{\pm}(\boldsymbol{\kappa}_t,t) - \dfrac{\rho_{\pm}(\boldsymbol{\kappa}_t,t)}{T_2} - i \chi_{\pm}(\boldsymbol{\kappa}_t,t) \big[ \rho_{+}(\boldsymbol{\kappa}_t,t) - \rho_{-}(\boldsymbol{\kappa}_t,t) \big].
    \end{array}
    \right.
\end{equation}
\end{widetext}
We define the instantaneous energy gap $\hbar \omega_{g}(\boldsymbol{\kappa}_t) = \epsilon_g(\boldsymbol{\kappa}_t) = \epsilon_+(\boldsymbol{\kappa}_t) - \epsilon_-(\boldsymbol{\kappa}_t)$ and the Rabi frequency associated to the interband coupling $\chi_{\pm}(\boldsymbol{\kappa}_t,t) = q_e \mathbf{E}(t) \cdot \boldsymbol{\mathcal{A}}_\pm(\boldsymbol{\kappa}_t)/\hbar$, proportional to the Berry connection $\boldsymbol{\mathcal{A}}_\pm(\mathbf{k}) =  i \langle \phi^+_{\mathbf{k}} | \nabla_{\mathbf{k}} | \phi^-_{\mathbf{k}} \rangle$. 
Electron decoherence times in graphene have been experimentally determined to be on the order of tens of femtoseconds \cite{Hommelhoff2021_T2_graphene}. It has been customary, however, to introduce unrealistically short decoherence times (around 1~fs) in theoretical calculations in order to smooth the harmonic spectra and recover harmonic structures resembling those observed in experiments. Spatial averaging of the current density over the beam profile at the target has pointed to provide such smoothing~\cite{Floss2018, Boyero2021_transverse_phase_matching}. In the non-averaged case---i.e., the microscopic emission from graphene---our calculations show that decoherence times larger than 10~fs have barely any effect on the harmonic spectrum, and are thus irrelevant for values as large as those determined experimentally. The harmonic emission is properly reproduced from  Eqs.~\eqref{SBE_bands_base} in the limit $T_2 \rightarrow \infty$.
The implications of using artificially short phenomenological dephasing times, and their connection with our study, are discussed in Sec. I of the Supplemental Material.

The matrix elements of the velocity operator read as
\begin{multline}\label{velocity_matrix_elements}
     \hat{\mathbf{v}}_{\mathbf{k}}^{nn'}  \equiv \langle \phi^n_{\mathbf{k}} | \hat{\mathbf{v}} | \phi^{n'}_{\mathbf{k}} \rangle =\\= \frac{1}{\hbar} \Big( \delta_{nn'} \nabla_{\mathbf{k}} \epsilon_n(\mathbf{k}) + i \pmb{\mathcal{A}}_{nn'}(\mathbf{k}) \big[ \epsilon_n(\mathbf{k}) - \epsilon_{n'}(\mathbf{k}) \big] \Big)
\end{multline}
for the bands indices $n,n' = \pm$. Note that the first term in Eq.~\eqref{velocity_matrix_elements} accounts for the intraband contribution, while the second term corresponds to the interband current. The total velocity of the charge carriers is then computed as the trace with the density matrix,
\begin{equation}\label{total_velocity}
    \mathbf{v}(t) = \int_{\mathrm{BZ}} \mathrm{Tr} \big[ \hat{\mathbf{v}}_{\boldsymbol{\kappa}_t} \rho _{\boldsymbol{\kappa}_t} \big] \mathrm{d}^2 \mathbf{k}, 
\end{equation}
where the integration extends to the SLG first Brillouin zone (BZ). 
The power of the electromagnetic radiation emmited by accelerated charges under the dipolar approximation and in the non-relativistic regime is given by the Larmor formula $ \propto | \mathbf{a}(t) |^2$, being $\mathbf{a}(t) = \dot{\mathbf{v}}  (t)$ the dipole acceleration. Thus, the microscopic harmonic signal is given by its Fourier transform, $\sim | \mathbf{a}(\omega) |^2$.

In this work, we consider a driving field with central wavelength of 3 $\mu$m and two driving peak intensities $I_0$ below the damage threshold of graphene \cite{Roberts2011_damage_threshold}. The field linear polarization is directed along the $\Gamma-K$ symetry axis of graphene. Figure~\ref{HHG_scheme}(a) shows the microscopic HHG spectra and the corresponding temporal profiles for $I_0 = 5 \times 10^{10}$~$\mathrm{W}/\mathrm{cm}^2$ (top panels) and $I_0 = 10^{12}$~$\mathrm{W}/\mathrm{cm}^2$ (bottom panels), obtained after high-pass filtering above harmonic orders 5th and 17th, respectively. The temporal envelope of the driving field is modelled as a sine-squared function with a total duration of 8 optical cycles, corresponding to a $28.8$~fs full width at half maximum (FWHM).

The calculation of the macroscopic emission follows the strategy developed for gas-phase HHG \cite{Carlos2010_macroscopic_HHG} adapted to SLG \cite{Boyero2021_transverse_phase_matching,GarciaCabrera2024}. The total electromagnetic field at the detector plane is computed as the coherent superposition of the far-field harmonic radiation emitted by point-like elementary sources at the target. 
The propagation to the far-field detector is obtained by solving the Maxwell's propagator, where the contribution of the $j$-th dipole at cell position $\mathbf{r}_j$ to a detector pixel located at $\mathbf{r}_d$ can be expressed as
\begin{equation}\label{jth_dipole_radiation}
    \mathbf{E}_j(\mathbf{r}_d,t) = \frac{q_j}{c^2 |\mathbf{r}_d - \mathbf{r}_j|} \, \mathbf{s}_d \times \big[  \mathbf{s}_d \times  \mathbf{a}_j (t_r) \big],
\end{equation}
where $\mathbf{s}_d$ is the unit vector pointing towards $\mathbf{r}_d$, and the dipole acceleration $\mathbf{a}_j$ is evaluated at the retarded time $t_r = t - \frac{|\mathbf{r}_d - \mathbf{r}_j|}{c}$. The arbitrarily large detector distance must fulfill $|\mathbf{r}_d| \gg |\mathbf{r}_j|$. The harmonic far-field is then computed as the coherent sum $\mathbf{E}(\mathbf{r},t) = \sum_{j=1}^N \mathbf{E}_j(\mathbf{r},t)$, where $N$ is large enough to ensure convergence. 

In this work, the Gaussian driving field exhibits a beam waist of 30~$\mu$m, with a corresponding Rayleigh distance of $z_R = 942$~$\mu$m, and we pay particular attention to the relative position between the SLG target and the focus of the driving beam, denoted by $z_0$ (see Fig.~\ref{HHG_scheme}).



\section{Results and Discussion}

\begin{figure*}[t]
\centering
\includegraphics[width=\textwidth]{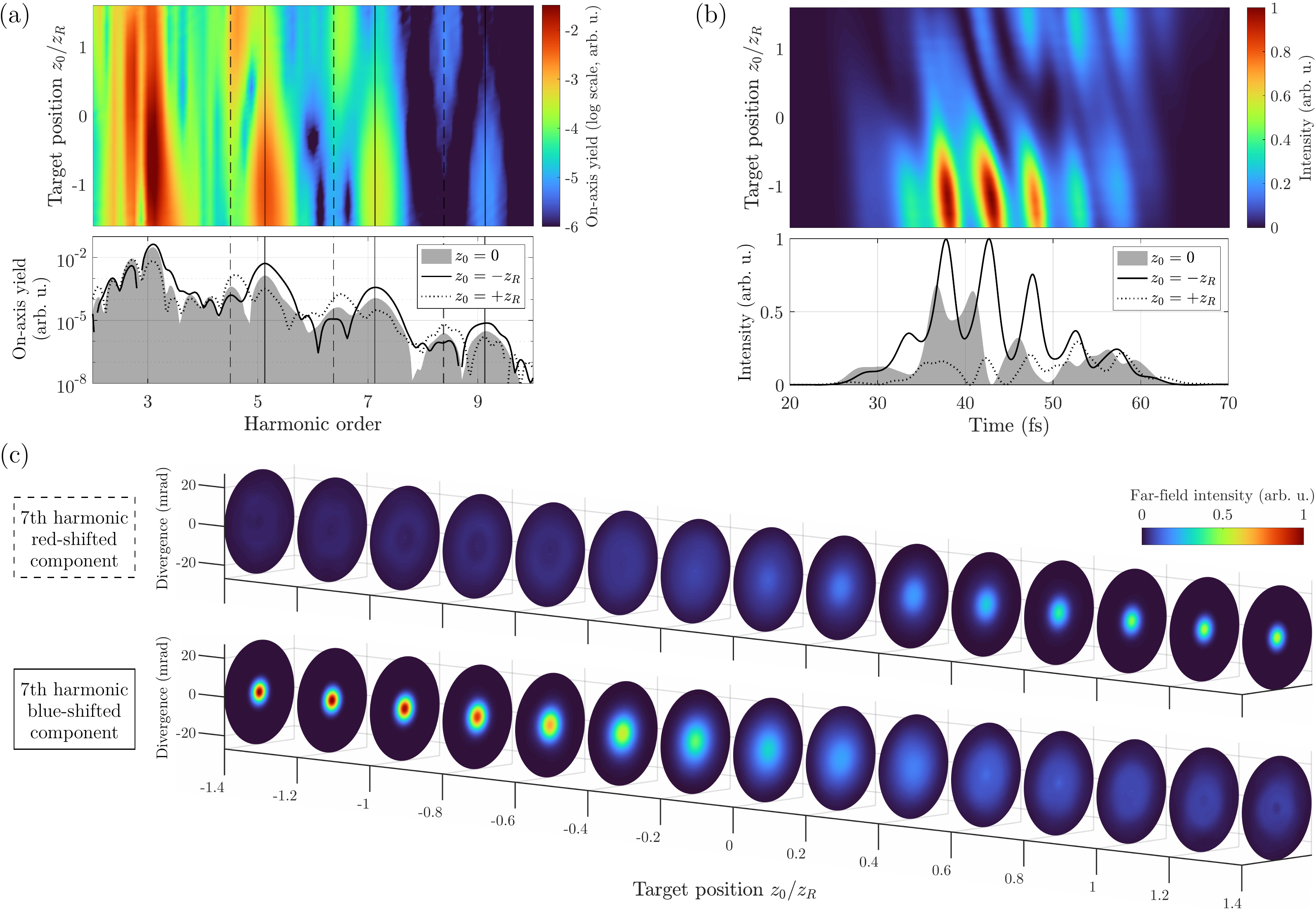}
\caption{Results for HHG in SLG driven by a low-intensity Gaussian beam with peak intensity $I_0 = 5 \times 10^{10}$~$\mathrm{W}/\mathrm{cm}^2$. (a) On-axis far-field harmonic yield depending on the target position $z_0$ with respect to the driving beam focus. HHG spectra for positions $z_0 = 0$ (gray shaded area), $z_0 = -z_R$ (solid line) and $z_0 = +z_R$ (dashed line) are highlighted in the bottom panel. Vertical solid (dashed) lines correspond to the blue-shifted (red-shifted) frequency components for three harmonics of interest (5h, 7th and 9th). (b) On-axis far-field temporal harmonic emission after performing the FT of results in (a), filtering from the 5th harmonic. Temporal profiles for positions $z_0 = 0$ (gray shaded area), $z_0 = -z_R$ (solid line) and $z_0 = +z_R$ (dashed line) are highlighted in the bottom panel. (c) Far-field intensity profiles of the red- (top) and blue-shifted (bottom) frequency components of the 7th harmonic for different target positions with respect to the driving beam focus.}
\label{results_low_intensity}
\end{figure*}

We first consider a low-intensity regime, $I_0 = 5 \times 10^{10}$~$\mathrm{W}/\mathrm{cm}^2$. Already at the microscopic level, Fig.~\ref{HHG_scheme}(a), the HHG spectrum exhibits a well-resolved double-peak structure within each harmonic order, which we shall refer to as red- and blue-shifted spectral components. Figure~\ref{results_low_intensity}(a) shows the macroscopic on-axis far-field HHG spectrum as a function of the relative position $z_0$ of the SLG target with repect to the driving beam focus, placed at $z=0$. We scale the driving field amplitude to obtain the same peak intensity at each target position. As a consequence, the maximum photon energy of the HHG spectrum remains invariant to $z_0$. Note, however, that the spectral shape changes with the target position, and, in particular, the relative weight of the two frequency components varies strongly as the SLG is displaced away from the focus. Indeed, the red-shifted contributions are found to be selectively suppressed (enhanced) compared to the blue-shifted ones when moving the SLG before (after) the focus. This behavior is highlighted in the bottom panel of Fig.~\ref{results_low_intensity}(a) by comparing the spectra obtained at $z_0=0$ (gray shaded area), $z_0=-z_R$ (solid line), and $z_0=+z_R$ (dashed line). The macroscopic response, therefore, produces a cleaner, more intense and more pronounced harmonic spectrum if the SLG is placed before the focus, where the red-shifted components are suppressed. A detailed analysis (see Supplemental Material Sec. I) demonstrates that the blue-/red-shifted frequency components originate predominantly from the emission of short-/long-time elctron-hole recombinations during the leading/trailing parts of the driving pulse, respectively. This attribution suggests that imperfect recollisions~\cite{Mette.B.Gaarde2020_imperfect_recollisions,Boyero2022_imperfect_recollisions} and recombination events involving large electron-hole separations~\cite{G.G.Brown2023_part1,G.G.Brown2023_part2} contribute mainly to the red-shifted components.

The selection of harmonic contributions when moving the SLG sheet with respect to the driving-beam focus not only cleans the spectrum, but also provides the conditions required for synthesizing regular pulse trains. Figure~\ref{results_low_intensity}(b) shows the temporal profile of the on-axis harmonic emission obtained by Fourier transforming the spectra shown in Fig.~\ref{results_low_intensity}(a) after filtering out harmonic orders below the fifth. Remarkably, when the SLG sheet is placed before the focus, the suppression of the red-shifted/long-time contributions results in the emission of a regular pulse waveform, sustained by a well-defined phase relationship among the harmonics. This is further illustrated in the bottom panel of Fig.~\ref{results_low_intensity}(b), where the temporal emission is compared for $z_0=0$ (gray shaded area), $z_0=-z_R$ (solid line), and $z_0=+z_R$ (dashed line). Note, however, that the femtosecond waveform is not yet structured into fully isolated attosecond bursts because of the absence of a broad and well-defined harmonic plateau. On the other hand, when the SLG sheet is placed after the focus, red-shifted/long-time contributions are enhanced. However, their lower efficiency and reduced contrast with respect to the blue-shifted/short-time contributions lead to a more irregular waveform. This behavior is reminiscent of the competition between short- and long-trajectory contributions in gas-phase HHG.

\begin{figure}
\centering
\includegraphics[width=\linewidth]{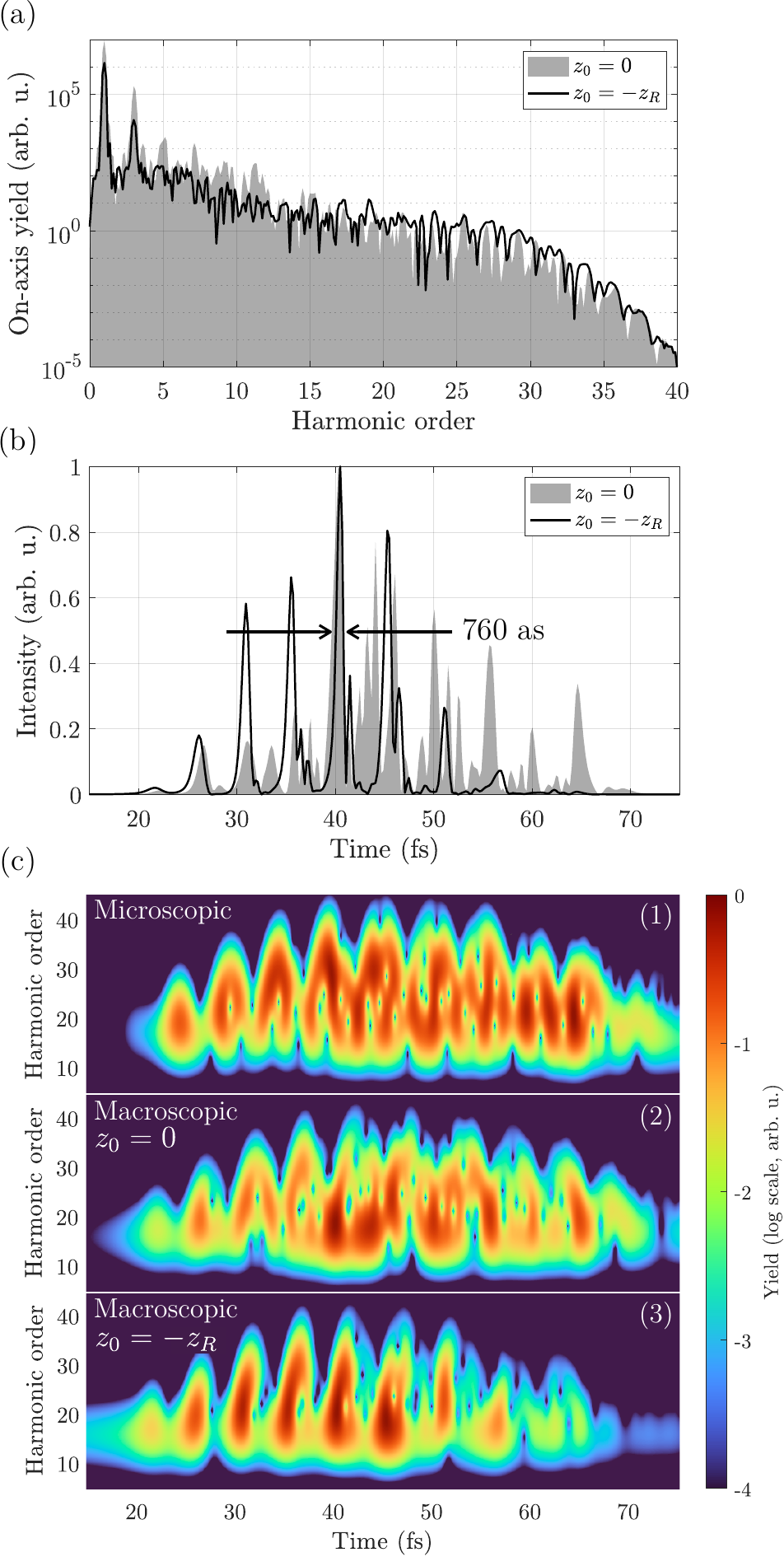}
\caption{Results for HHG in SLG driven by a high-intensity Gaussian beam with peak intensity of $I_0 = 10^{12}$~$\mathrm{W}/\mathrm{cm}^2$. (a) Comparison between the on-axis far-field harmonic yield with the SLG target placed at the beam waist $z_0 = 0$ (gray shaded area) and before the focus $z_0 = -z_R$ (solid line). (b) On-axis far-field temporal harmonic emission after performing the FT of results in (a), filtering from the 17th harmonic. When placing the SLG before the focus (solid line), a regular attosecond pulse train is detected. (c) Time-frequency analysis of the microscopic HHG response (1) and the on-axis far-field harmonic signal when the target is placed at $z_0 = 0$ (2) and $z_0 = -z_R$ (3). Whereas in (1) and (2) the spectrogram presents irregular patterns associated to multiple electronic contributions, in the later case a regular positive-slope structure enables the synthesis of regular positively-chirp attosecond pulse trains.}
\label{results_high_intensity}
\end{figure}

The on-axis far-field intensity ratio between the blue- and red-shifted harmonic contributions differs because of their distinct divergence properties. This is shown in Fig.~\ref{results_low_intensity}(c), where we plot, for the seventh harmonic, the far-field intensity profiles of the red-shifted (top) and blue-shifted (bottom) contributions as a function of the SLG position $z_0$. Before the focus ($z_0<0$), the blue-shifted components propagate more collimated, whereas the red-shifted contributions form more divergent harmonic beamlets. After the focus ($z_0>0$), although the red-shifted components become more collimated, their relative weight remains comparable to that of the blue-shifted contributions. These results demonstrate that the harmonic wavefront generated during HHG can be compensated by the driving-beam wavefront differently for the red- and blue-shifted frequency components, in close analogy with short- and long-trajectory selection in gas-phase HHG~\cite{wikmark2019spatiotemporal,Quintard2019,MartinHernandez2026}. In particular, placing the SLG sheet before the focus, at $z_0=-z_R$, provides the strongest compensation and spectral selection, yielding the smallest---least divergent---far-field harmonic beamlets, corresponding with the blue-shifted contributions.

To achieve an extended harmonic plateau that can support attosecond pulse trains, we next consider a higher driving peak intensity of $I_0 = 10^{12}$~$\mathrm{W}/\mathrm{cm}^2$, still well below the graphene damage threshold~\cite{Roberts2011_damage_threshold}. Figure~\ref{results_high_intensity}(a) shows the numerical results for the macroscopic on-axis HHG spectrum when the SLG is placed at the focus, $z_0=0$ (gray shaded area), and before the focus, at $z_0=-z_R$ (solid line). Again, the peak intensity has been scaled to achieve the same extension of the harmonic plateau. Both spectra are normalized relative to the value of their cut-off harmonic for comparison purposes. At this higher driving intensity, the HHG spectrum presents a clear non-perturbative plateau extending up to approximately the 29th harmonic order. As in the lower-intensity regime, placing the SLG before the focus leads to a cleaner harmonic spectrum. However, in contrast to Fig.~\ref{results_low_intensity}(a), the HHG spectra exhibit a complex structure due to the larger number of electron-hole trajectories contributing to each harmonic order, together with stronger interference between intraband and interband currents. This increased complexity blurs the distinction between the blue- and red-shifted contributions within each harmonic, although the same macroscopic cleaning effect remains. After Fourier transforming the HHG spectra and filtering out harmonic orders below the 17th, we obtain the temporal emission shown in Fig.~\ref{results_high_intensity}(b). When the SLG is placed before the focus (solid line), a clean and regular train of attosecond bursts with a pulse duration of $760$~as FWHM is obtained, comparable to those generated in gas-phase HHG. This temporal structure holds for higher far-field divergences, where the spatial filtering is evident (see Supplemental Material Sec. II).

The clean attosecond emission arises from the regular phase relationship established among the plateau harmonics after phase-matching filtering in the macroscopic propagation. This interpretation is further supported by the time-frequency analysis (TFA) shown in Fig.~\ref{results_high_intensity}(c). Fig.~\ref{results_high_intensity}(c.1) shows the TFA of the microscopic emission, while the spectrograms in Figs.~\ref{results_high_intensity}(c.2) and \ref{results_high_intensity}(c.3) correspond to the macroscopic emission obtained when the SLG is placed at $z_0=0$ and $z_0=-z_R$, respectively---thus obtained from the temporal emissions in Fig.~\ref{results_high_intensity}(b). In all three cases a Gaussian window of 1.2~fs FWHM has been used to perform the Fourier transform of each Gaussian-masked emission. The spectrograms reveal that, under appropriate macroscopic conditions, the harmonic emission can be effectively cleaned, leading to a regular phase relationship among the contributing harmonics. Although the red- and blue-shifted components cannot be clearly resolved at this higher intensity, the TFA in Fig.~\ref{results_high_intensity}(c.3) clearly shows a strong suppression of the emission from the trailing part of the driving pulse, associated with long-time contributions, i.e. delayed electron-hole recombinations or imperfect recollisions. This behavior can be then attributed to the distinct macroscopic propagation of these contributions, which leads to different divergence and, consequently, to their selective suppression when optimal wavefront compensation between the harmonic field and the driving beam is achieved. In contrast to gas-phase HHG, distinguishing between short- and long-trajectory contributions is less straightforward due to the increased complexity of electronic pathways in solids. Nevertheless, the time-frequency analysis reveals an analogous behavior: long-time contributions, linked to emission from the trailing part of the driving pulse, are suppressed under favorable phase-matching conditions when the target is placed before the focus. This enables the synthesis of attosecond pulse trains with a positive chirp---or attochirp---, as evidenced by the positive slope of the spectrogram features in Fig.~\ref{results_high_intensity}(c.3), a hallmark of short-trajectory-dominated emission in phase-matched gas-phase HHG.

Our numerical simulations therefore reveal that macroscopic phase matching and propagation in solid-state HHG not only clean the harmonic spectrum, but also enable the synthesis of positively chirped attosecond pulse trains, in close analogy with gas-phase HHG.
We note that previous studies \cite{Vampa2014,ZhaopinChen2025,Venkat2026} report attosecond pulse generation in solids from calculations using unrealistic electron dephasing times to clean the HHG spectrum. Noteworthly, we demonstrate, instead, that the cleaner experimental spectra---and corresponding attosecond pulse train---find a physical origin in the spectral suppression of phase mismatched constributions. As shown in Sec. I of the Supplemental Material, introducing short phenomenological dephasing times produces a similar cleaning of the harmonic spectrum. In addition, our findings support recent theoretical works showing that longer electron-hole recombination pathways can be suppressed at the macroscopic level~\cite{G.G.Brown2023_part1,G.G.Brown2023_part2}. Here, we show that this suppression is effectively achieved when suitable transverse phase-matching conditions are established, and that it is crucial for achieving attosecond pulse synthesis in experiments of solid-state HHG.

Beyond providing a realistic framework for attosecond pulse synthesis in solid-state HHG, our results identify a practical control knob by tailoring of the driving-beam wavefront. While we have demonstrated this control at the generation stage, by positioning the SLG target with respect to the driving focus, an analogous selection can also be achieved after generation by refocusing the harmonic emission. In Sec. III of the Supplemental Material, using Fresnel propagation, we show that the blue-shifted contributions associated with short-time contributions in the leading part of the pulse can be selected by appropriate refocusing of the harmonic beam.

\section{Conclusion}

Our results demonstrate that transverse phase-matching in solid-state HHG enables the synthesis of attosecond pulse trains. Short-time recombination events can be selected through an appropriate wavefront configuration of the driving beam, in close analogy with the selection of short-trajectory contributions in gas-phase HHG. Conversely, longer-time recombination events, such as imperfect recollisions or recombinations involving large electron-hole separations, can be suppressed during macroscopic propagation due to their less favorable phase relationship and larger accumulated phase. As a result, we have shown that, in SLG, placing the target at an optimal position before the focus leads to the generation of a clean, regular, positively chirped attosecond pulse train. A similar selection can also be achieved through suitable refocusing of the harmonic emission.

Although our study focuses on SLG, we anticipate that HHG in other gapless or finite-gap two-dimensional/thin materials may also benefit from this macroscopic selection mechanism, enabling cleaner temporal emission approaching the attosecond regime. Since, up to now, attosecond pulses have been only demonstrated in tabletop gas-phase HHG and in free-electron lasers \cite{Franz2024, Driver2024}, our results demonstrate a pathway for the development of brighter compact attosecond light sources presenting the full spatial and temporal coherence of HHG sources, making it invaluable for applications in imaging, metrology or spin/charge dynamics with nanometer and attosecond resolutions.

In addition, one of the key strengths of HHG in solids is its sensitivity to the underlying electronic dynamics, including interband and intraband contributions. However, mapping the measured harmonic emission back onto the microscopic electronic currents is highly nontrivial, since the macroscopic signal is shaped not only by the local nonlinear response, but also by propagation and phase-matching effects. Therefore, a quantitative reconstruction of electronic currents requires a proper understanding of how these macroscopic effects modify the emitted harmonics. In this sense, our work not only opens a route towards the generation of solid-state attosecond pulses, but also provides a framework to disentangle electronic dynamics exploiting phase-matching mechanisms in solids.


\begin{acknowledgments}
We acknowledge funding from Ministerio de Ciencia e Innovaci\'on (Grant PID2022-142340NB-I00) and the Department of Education of the Junta de Castilla y Le\'on and FEDER Funds (No. SA108P24 and Escalera de Excelencia CLU-2023-1-02). This project has received funding from the European Research Council (ERC) under the European Union’s Horizon Europe research and innovation programme (ERC Proof of Concept grant agreement No. 101247206).
\end{acknowledgments}


\section*{SUPPLEMENTAL MATERIAL}

\subsection*{I. Identification of blue-/red-shifted spectral componentswith short-/long-time contributions}

To elucidate the origin of the spectral characteristics shown in the main text, we perform here a detailed microscopic study of HHG in SLG. In the context of HHG in finite-gap solids, some studies address the role of the spectral interference between short and long trajectories, and its connection with the use of artificially short dephasing times \cite{Y.W.Kim2019_spectral_interference_solids, T.Y.Du2019_spectral_properties_solids}. However, in gapless materials such as SLG, the distinction between these quantum paths is not so trivial as pairs are created constantly during the interaction instead of at the maxima of the external field. In order to temporally discriminate the electron-hole recombination events we have performed a spectrographic analysis of the microscopic dipolar emission. First, we define a temporal filter mask $W(t,t_0)$ as a unitary step function centered at time $t_0$ with a smooth stepness of width $2 \sigma$,
\begin{equation}\label{off_mask}
    W(t,t_0) := 
    \left\{
    \begin{array}{cc}
    1 & t \leq t_0 - \sigma \\
    \cos^2 \left[ \dfrac{\pi}{2} \dfrac{ t - (t_0 - \sigma)}{2\sigma} \right] & | t - t_0 | < \sigma \\
    0  & t \geq t_0 + \sigma
    \end{array}
    \right. .
\end{equation}
\begin{figure}[ht]
\centering
\includegraphics[width=\linewidth]{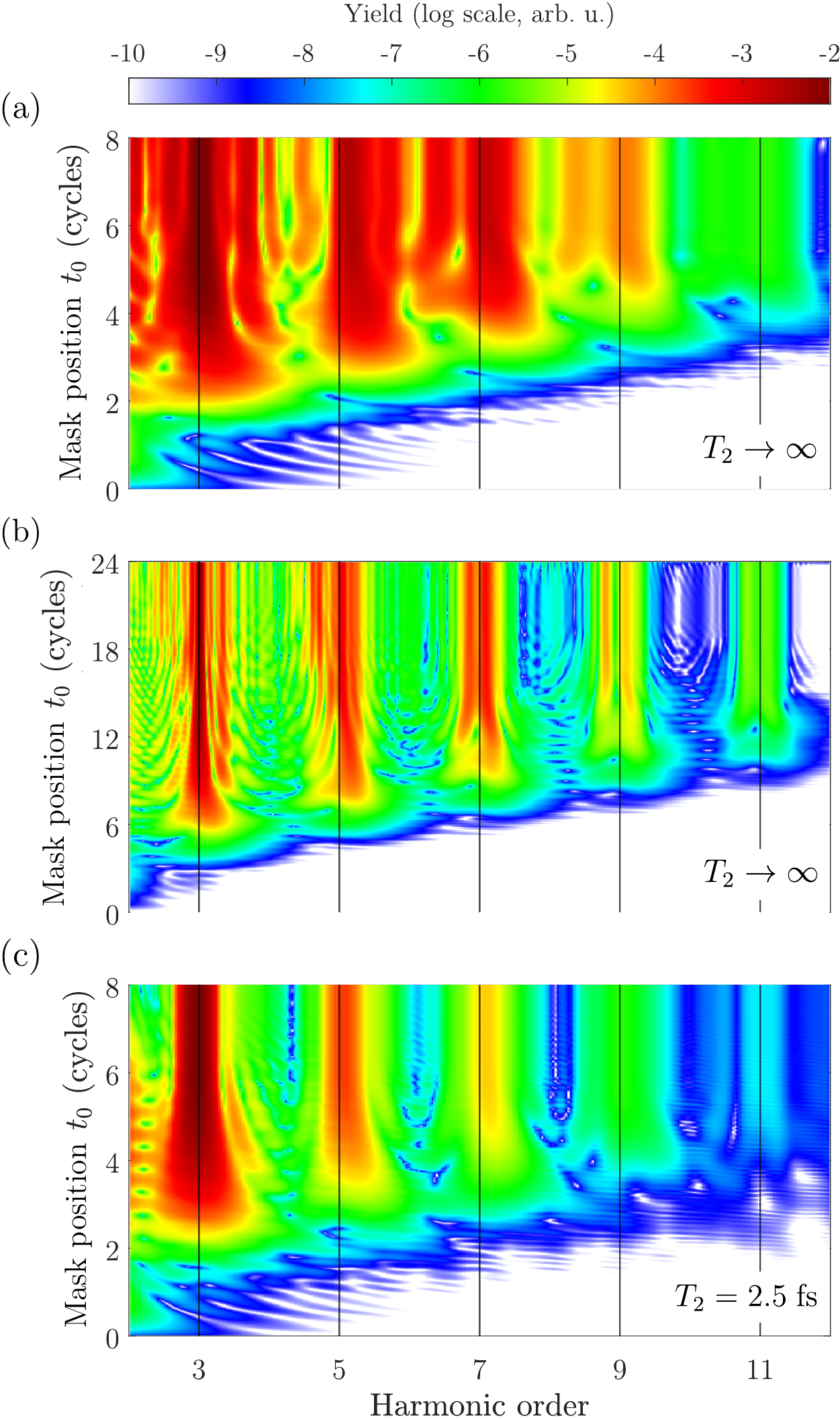}
\caption{Spectrograms of the microscopic HHG emission masked with the window function $W(t,t_0)$ of width $2 \sigma = 10$~fs centered at time $t_0$, using a mid-IR 3~$\mu$m-wavelength driving field with peak intensity of $I_0 = 5 \times 10^{10}$~$\mathrm{W}/\mathrm{cm}^2$, using (a) a short (8 cycles total duration, 28.8~fs FWHM) driving pulse, as in the main text, and (b) a longer driving pulse of 24 cycles total duration, 86.4~fs FWHM. In panel (c) an artificial dephasing time of $T_2= 2.5$~fs is included for the first case of 28.8 fs FWHM.}
\label{microscopic_results}
\end{figure}

We present in Figure~\ref{microscopic_results} the spectrograms of the microscopic emission dipole in the low peak intensity case of $I_0 = 5 \times 10^{10}$~$\mathrm{W}/\mathrm{cm}^2$. We set a stepness of one driver cycle, $2\sigma = 10$~fs, and scan the parameter $t_0$. The double-peak structure in the spectum arises approximatelly after half of the interaction with the driver, see Fig.~\ref{microscopic_results}(a). Then, we can distinguish between short-time electron-hole recombination events in the leading part of the laser pulse generating blue-shifted frequency components for each harmonic, and long-time events in the trialing part of the laser pulse contributing to the total yield with red-shifted frequency components---the latter including imperfect recollisions that take place at longer interaction times \cite{Mette.B.Gaarde2020_imperfect_recollisions,Boyero2022_imperfect_recollisions}. These frequency shifts allow to discern between time events before (increasing intensity) and after the pulse peak (decrasing intensity).

To further justify the physical origin of the shown spectral structure, we present in Fig.~\ref{microscopic_results}(b) the spectrogram of the microscopic dipole driven by a slowly-varying envelope pulse with a longer duration of 24 cycles, corresponding to 86.4~fs FWHM. In this case, frequency shifts are less evident, making the double-peaks closer, and resulting in sharper harmonics with less broadened spectral content. 

The red-shifted components attributed to long-time recombination events can be indeed suppressed by considering an artificially short dephasing time $T_2$ in the SBE. Fig.~\ref{microscopic_results}(c) shows the spectrogram computed in the same conditions as in  Fig.~\ref{microscopic_results}(a), adding a dephasing time of a quarter cycle, $T_2= 2.5$~fs. In this case harmonics are much better defined with a single-peak structure, as reported in several theory works of solid-state HHG \cite{Vampa2014,ZhaopinChen2025,Venkat2026}. This analysis thus demonstrates the identification between blue-/red-shifted and short-/long-time contributions to the HHG spectrum, and the connection of our work with the inclusion of artificially short dephasing times.


\subsection*{II. Attosecond pulse emission at higher far-field divergences}

\begin{figure}[ht]
\centering
\includegraphics[width=\linewidth]{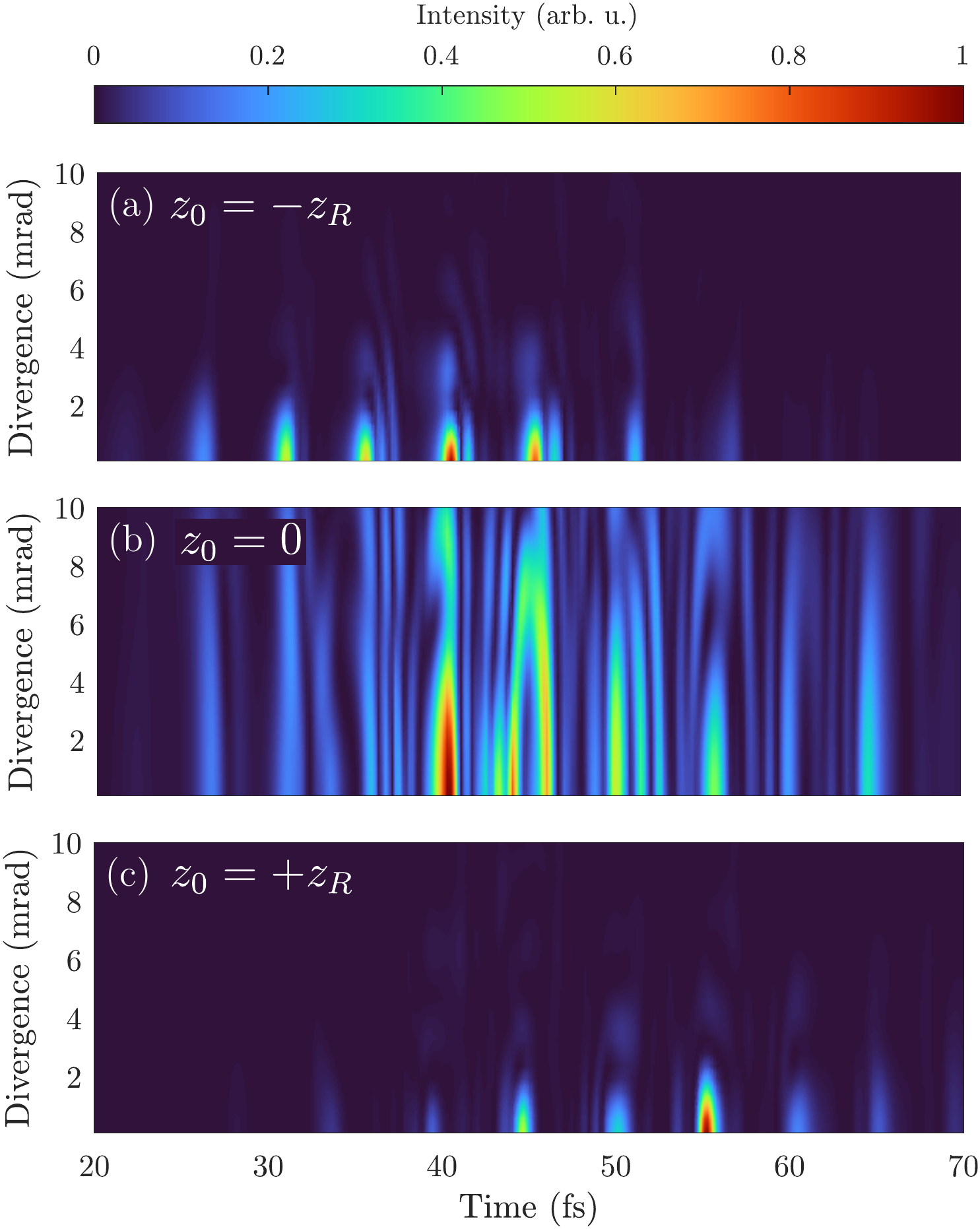}
\caption{Spatial far-field profile of the temporal emission of high harmonics driven by a Gaussian beam of peak intensity $I_0 = 10^{12}$~$\mathrm{W}/\mathrm{cm}^2$ when the SLG sheet is placed at (a) $z_0 = -z_R$, (b) $z_0 = 0$, and (c) $z_0 = -z_R$.}
\label{results_off-axis}
\end{figure}

In the main text, spectra and temporal emission of harmonics have been shown when detected on-axis at the spatial point of maximum intensity. In order to provide for a complete analysis that can be compared against experimental realizations we show here the complete spatial profile of the far-field attosecond pulses, for the same driving field parameters as in the main text. When the SLG is placed at the driving beam focus, $z_0 = 0$, the irregular spatial profile hinders the generation of an attosecod pulse beam, see Fig.~\ref{results_off-axis}(b). However, when the SLG is placed out of focus, a clean spatial attosecond pulse train profile is obtained. When the SLG plane is placed before/after the driving beam focus at $z_0 = \mp z_R$, Figs.~\ref{results_off-axis}(a) and (c) respectively, lower-/higher- diverging short-/long-time recombinations are present. This selection allows for the generation of a far-field attosecond pulse train beam that is more intense and cleaner when the SLG is placed before the focus; i.e. when short-time contributions are selected. Note that the direct comparison between panels (a) and (c) allows to further identify short-/long-time contributions.  

\subsection*{III. Re-focusing of far-field high-order harmonics and attosecond pulse trains}

\begin{figure}
\centering
\includegraphics[width=\linewidth]{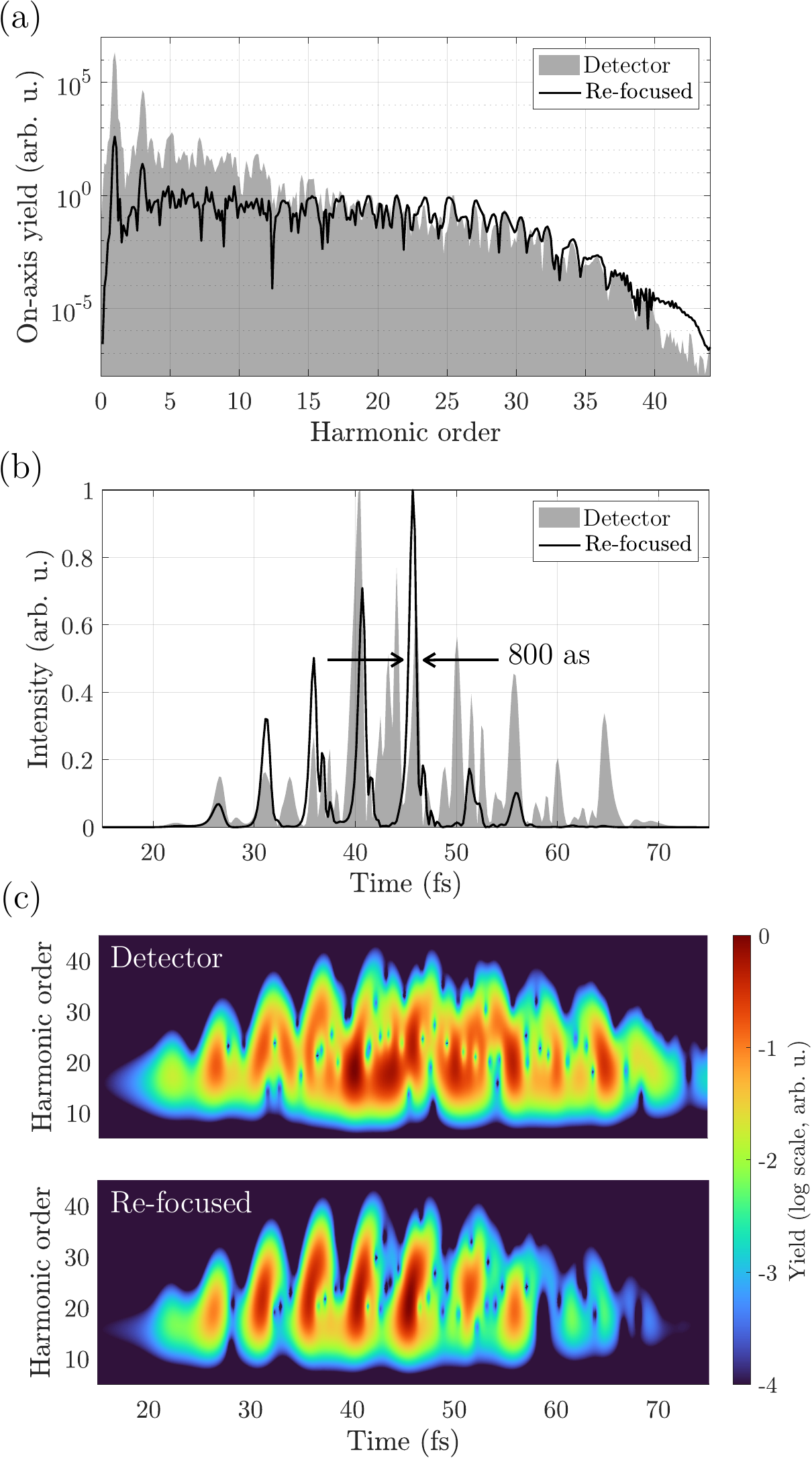}
\caption{Results for HHG in SLG driven by a Gaussian beam of peak intensity  $I_0 = 10^{12}$~$\mathrm{W}/\mathrm{cm}^2$ with the SLG target placed at the focus position. In panels (a) and (b) we show the comparison between the HHG spectra and temporal emission (filtering from the 17th harmonic) between the on-axis far-field signal recorded at the detector plane (gray shaded area) and the re-focused emision after free propagation over a distance of $z' = 5$~cm (solid line). (c) Time-frequency analysis for the signal recorded at the detector (top) and the re-focused emision (bottom).}
\label{results_refocusing}
\end{figure}

In the main text, we have shown that clean far-field attosecond pulse trains can be obtained by adjusting the wavefront curvature of the Gaussian driving beam through the relative position between the SLG sheet and the focus. However, similar control can also be achieved by exploiting the re-focusing properties of the emitted harmonics due to the different divergence properties of the blue- and red-shifted frequency components associated to short- and long-time events in electron-hole recombinations, respectively. 

We perform macroscopic HHG simulations in SLG placed at the focus of the Gaussian driving beam with a waist of 30~$\mu$m. The driving field parameters are as those presented in the main text: linearly polarized, mid-IR 3~$\mu$m-wavelength, $28.8$~fs FWHM and peak intensity of $I_0 = 10^{12}$~$\mathrm{W}/\mathrm{cm}^2$. Once the harmonic profiles are computed at the far-field detector, we introduce a parabolic phase ($f'=5$ cm) to focus HHG radiation. We compute the subsequent propagation over a distance $z'$ solving the Fresnel diffraction integral. Figure~\ref{results_refocusing}(a) shows the on-axis spectrum recorded at the far-field detector (gray shaded area) and the cleaned re-focused harmonic yield after propagation over a distance of $z'= f' = 5$~cm (solid line). Both spectra are normalized relative to the value of their cut-off harmonic for comparison. The temporal emission for both cases are presented in Fig.~\ref{results_refocusing}(b) afer filtering out harmonics below the 17th. When harmonics are re-focused, a regular train of attosecond pulses is obtained (solid line). This distance value has found to be the optimum for the highest compression up to 800 as FWHM, obtaining attosecond pulses with a quality comparable to those generated with the target plane before the driver focus (main text). The underlaying mechanism remains the same: long-time contributions present higher far-field divergence when the SLG is palced at the focus, and, upon refocusing, the resulting HHG spectra is dominated by short-time contributions, enabling the generation of clean and regular attosecond pulse trains. This analysis is further reinforced by the corresponding time-frequency analyses presented in Fig.~\ref{results_refocusing}(c), for the far-field (top) and refocused (bottom) emission.


\bibliography{bibliography}




\end{document}